\documentclass{PoS}

\newcommand{\be}{\begin{equation}}
\newcommand{\ee}{\end{equation}}
\newcommand{\ba}{\begin{eqnarray}}
\newcommand{\ea}{\end{eqnarray}}
\renewcommand{\(}{\left(}
\renewcommand{\)}{\right)}

\newcommand{\lk}{\left[}
\newcommand{\rk}{\right]}

\newcommand{\w}{\omega}

\title{Holographic thermalization patterns}

\ShortTitle{Holographic thermalization patterns}

\author{\speaker{Stefan STRICKER} \\  
     Institute for Theoretical Physics, Vienna University of Technology\\
      Wiedner Hauptstr. 8-10/136, 1040 Vienna, Austria\\
     
       E-mail: \email{stricker@hep.itp.tuwien.ac.at}}

\abstract{We investigate the behaviour of various correlators in $\mathcal{N}=4$ super Yang Mills theory, taking finite coupling corrections into account. In the thermal limit we investigate the flow of the quasinormal modes as a function of the 't Hooft coupling. Then by using a specific model of holographic thermalization we investigate the  deviation of the spectral densities from their thermal limit in an out-of-equilibrium situation.
The main focus lies on the thermalization pattern with which the various plasma constituents of different energies approach their final thermal distribution as the coupling constant decreases from the infinite coupling limit.
All results point towards the weakening of the usual top down thermalization pattern.}

\FullConference{The European Physical Society Conference on High Energy Physics \\
		 18-24 July, 2013\\
		 Stockholm, Sweden}

\begin{document}

\section{Introduction}
Understanding the processes behind the complicated field dynamics  in a relativistic heavy ion collision presents a  complicated  challenge to QCD theorists.
Experiments at RHIC and the LHC, together with the early onset of hydrodynamical behaviour, indicate that the produced state of matter behaves as a strongly coupled ideal fluid with very low shear viscosity over entropy ratio.
The strongly coupled nature of the quark gluon plasma (QGP) has led the gauge/gravity duality to become one of the standard tools to investigate its properties \cite{DeWolfe:2013cua, CasalderreySolana:2011us}.
The duality has proven particularly useful in investigating time dependent out-of-equilibrium phenomena and the approach to thermal equilibrium by mapping the strongly coupled field dynamics to black hole formation in asymptotically anti-deSitter (AdS) space time in one higher dimension.
This has led to the insight that the early applicability of hydrodynamics, sometimes called  hydroization or hydrodynamization, does not mean that the system is isotropic, let alone thermal. 

Another important question in a thermalizing system concerns the thermalization pattern with which the plasma constituents of different energies approach their final thermal distribution.
On the weakly coupled side classical calculations have shown  that the  thermalization process is of the bottom-up type, i.e. low energetic modes reach their thermal distribution first, with inelastic  scattering processes being the driving mechanism behind it \cite{Baier:2000sb}. 
In the early stages of the collision many soft gluons are emitted  which form a thermal bath rather quickly, which  then draws energy from the hard modes.
Recently this picture got supported by numerical simulations \cite{Berges:2013eia}. In \cite{Kurkela:2011ti} an alternative proposal was made: the thermalization process is driven by instabilities which isotropize the momentum distributions more rapidly than scattering processes.
On the contrary,
 holographic calculations in the infinite coupling limit always point towards top-down thermalization where the high energetic modes reach thermal equilibrium first, indicating a probable transition at intermediate coupling.
%Using geometric probes such as entaglemant entropy or Wilson loops this follows from a simple causal argument.

In this proceedings we are going to investigate the  thermalization patterns in $\mathcal{N}=4$ super-Yang Mills  (SYM)  plasma and its dependence on the coupling constant.
This is done through the determination of the off-equilibrium  retarded Greens functions whose evolution  gives direct information about how plasma  constituents of different energies approach their thermal distribution.
On the one hand we will investigate the plasma constituents themselves by looking at energy momentum tensor correlators \cite{Stricker:2013lma}.
On the other hand we will also use photons as probes for the thermalization process, which provide an observational window into the thermalization process, because once emitted they stream through the plasma almost unaltered \cite{Baier:2012ax, Baier:2012tc, Steineder:2012si, Steineder:2013ana}.
%The coupling constant dependence on the thermalization process using geometric probes was considered in \cite{Baron:2013cya}.

\section{The setup}
\subsection{the collapsing shell model}
We work within one of the simplest models of holographic thermalization, using the radial collapse of a spherical  shell of matter and the subsequent formation of a black hole \cite{Danielsson:1999zt}. 
Due to Birkhoff's theorem, outside the shell the metric is given by a black hole solution, whereas inside it is given by its zero temperature counterpart
\be\label{AdS5}
ds^2=\frac{r_h^2}{L^2 u}\lk f(u)dt^2+dx^2+dy^2+dz^2\rk+\frac{L^2}{4 u^2f(u)}du^2\;,
\ee
where
\be\label{f}
f(u) \,=\, \left\{ \begin{array}{lr}
f_+(u)=1- u^2 \, ,& \mathrm{for}\;u >u_s\\
f_-(u)=1\, ,& \mathrm{for}\; u < u_s
\end{array}\right. \;,
\ee
and $u\equiv r_h^2/r^2$ is a dimensionless coordinate where the boundary is located at $u=0$ and the horizon at $u=1$. From now on the index '--' denotes the inside and '+' the outside 
 space-time of the shell and  we set the curvature radius of the $AdS$ space to $L=1$. 

The shell can be described by the action for a membrane 
\be
S_m=-p \int d^4\sigma \sqrt{-\det g_{ij}}\;,
\ee
where $g_{ij}$ is the induced metric on the brane and $p$ is the only parameter that characterizes the membrane.
Due to the discontinuity of the time coordinate, fields living in the above background have to be matched across the shell using the Israel matching conditions given by 
\be
[K_{ij}]=\frac{\kappa_5^2 p}{3} g_{ij}\;,
\ee
where $[K_{ij}]=K^+_{ij}-K^-_{ij}$ is the extrinsic curvature and $\kappa_5$ denotes the gravitational constant. From the above equation   the trajectory of the shell is determined. We however are not going to treat the full dynamical process but
 work in the quasi static approximation, where the motion of the shell is slow compared to the other scales of interest and take snapshots  at different positions of the shell \footnote{Note that the quasi static approximation is not applicable at the latest stages of the collapse \cite{Lin:2013sga}. }.
 In the quasi static approximation the matching conditions  for the different fields  $\phi_i$
  we are considering simplify considerably and can be written in the uniform way \cite{Lin:2008rw}
  \be\label{mc}
\phi_i^-|_{u_s}=\sqrt{f_s}\phi_i^+|_{u_s}, \qquad \partial_u \phi_i^-+\frac{\kappa_5^2p}{3u}\phi_i^-|_{u_s}=f_s \partial_u \phi_i^+|_{u_s}\;.
\ee 
In what follows we set   $\kappa_5^2p=1$, which was set to zero in previous studies for photons \cite{Baier:2012ax, Baier:2012tc, Steineder:2012si, Steineder:2013ana}.

%Initial conditions of the shell  relevant for heavy ion collisions are determined through the relation  of the holographic coordinate with the temperature $r_h=T \pi$ and the saturation scale $r_s=Q_s\pi$ together with a vanishing initial velocity.
%For  the LHC these values are $T\sim 400$ MeV and $Q_s\sim1.23$ GeV.  
\subsection{Finite coupling corrections}

In order to go beyond the usual $\lambda=\infty$ limit  the type IIB supergravity action has to be supplemented by the first order string corrections of order $\alpha'^3$ in the inverse string tension \cite{Paulos:2008tn}
\be\label{action}
S_\mathrm{IIB}=\frac{1}{2\kappa_{10}}\int d^{10}x\sqrt{-g}\lk R_{10}-\frac{1}{2}(\partial \varphi)^2-\frac{1}{4.5!}(F_5)^2 \, +\gamma \;e^{\frac{-3}{2}}\varphi \Big(C+\mathcal{T}\Big)^4\rk\;.
\ee
$R_{10}$ denotes the Ricci scalar, $\gamma\equiv\frac{1}{8}\zeta(3)\lambda^{-\frac{3}{2}}$,
 $\varphi$ the dilaton field and $F_5$ the five-form field strength, while $C$ stands for the Weyl tensor and the tensor ${\mathcal T}$ is given by
\be
\mathcal{T}_{abcdef}=i\nabla_a F^+_{bcdef}+\frac{1}{16}\( F^+_{abcmn}F_{def}^{+\;\;mn}-3F^+_{abfmn}F_{dec}^{+\;\;mn}\) \, .
\ee
The index triplets $\{a,b,c\}$ and $\{d,e,f\}$ are understood to be first antisymmetrized with respect to all permutations, and the two triplets then symmetrized with respect to the interchange $abc\leftrightarrow def$. One should also note that our notation for the various contractions in equ.~(\ref{action}) is only schematic, and e.g.~the $C^4$ term in fact denotes the combination
\be
C_{hmnk}C_{pmnq}C_h^{\;rsp}C^q_{\;rsk}+\frac{1}{2}C^{hkmn}C_{pqmn}C_h^{\;rsp}C^q_{\;rsk} \, . \label{C4}
\ee 
For further details of this construction, see e.g.~\cite{Hassanain:2012uj} and references therein.

From the above action the $\gamma$-corrected metric is obtained
from which one can then obtain the $\gamma$-corrected  equations of motion (EoM) for a transverse U(1) vector field $E$ and metric perturbations of the form  $g_{\mu\nu}\rightarrow g_{\mu\nu}+ h_{\mu\nu}$.

Following  \cite{Kovtun:2005ev} the metric perturbations can be combined into  three gauge invariant fields $Z_i$ representing the three symmetry channels, namely the spin 0 (sound channel), spin 1 (shear channel) and the spin 2 (scalar channel). 
The EoMs are solved using standard AdS/CFT techniques. However, due to the presence of the shell the outside solution is a linear combination of the ingoing and outgoing modes, ($\phi_a={E,Z_i}$),
\be
\phi_a^+=c_+\phi_{a,in}+c_-\phi_{a,out}.
\ee
where the  coefficients $c_\pm$ are solved from the matching conditions (\ref{mc}). The spectral densities of the electric field and the gauge invariants are obtained  from
%\begin{subequations}
\be
\chi_E(\w,q,u_s,\gamma)=\frac{N_c^2 T^2}{2}\mathrm{Im}\lk \frac{E'_{+}(u)}{E_{+}(u)}\rk ,\qquad \chi_{Z_i}(\w,q,u_s,\gamma)=\frac{N_c^2 T^4}{2}\mathrm{Im}\lk \frac{Z''_{i,+}(u)}{ Z_{i,+}(u)}\rk 
\Bigg|_{u=0}
\ee
which need to be expanded to linear order in $\gamma$. For the thermalization processes it is useful to study  the relative deviation of the spectral densities from their thermal limit (denoted by 'th' below)
\be
\label{R}
R_a(\w,q,u_s\gamma)=\frac{\chi_a(\w,q,u_s\gamma)-\chi_\mathrm{a,th}(\w,q,\gamma)}{\chi_{a,\mathrm{th}}(\w,q,\gamma)}\, .
\ee

\section{Results}
Having outlined the main steps of our computation, we will now present the results, which are  divided into two parts. We start from the thermal limit and  inspect the effect of infinitesimal perturbations on the system by solving the  quasinormal mode (QNM) spectrum and by analyzing  the flow of the poles as functions of $\lambda$.
Then we use the collapsing shell model to study the behaviour of various spectral densities and their approach to thermal equilibrium taking the leading order string corrections into account. In the case of photons we also study the virtuality dependence.
\subsection{Quasi normal modes}
Quasinormal modes characterize the response of the system to infinitesimal  external perturbations and are the strong coupling equivalent to the quasiparticle picture of the weakly coupled field theory.
 They are solutions to linearized fluctuations of some bulk field obeying  incoming boundary conditions at the horizon and Dirichlet boundary conditions at the boundary. They appear as poles of the  corresponding retarded Green's function and have the generic form 
\be\label{frobenius}
\w_n(q)=M_n(q)-i\Gamma_n(q),
\ee 
 where $q$ is  the three momentum of the mode, $M_n$ and $\Gamma_n$ correspond to the mass and the decay rate of the excitation, respectively. 

\begin{figure}[h]
\centering
\includegraphics[width=7cm]{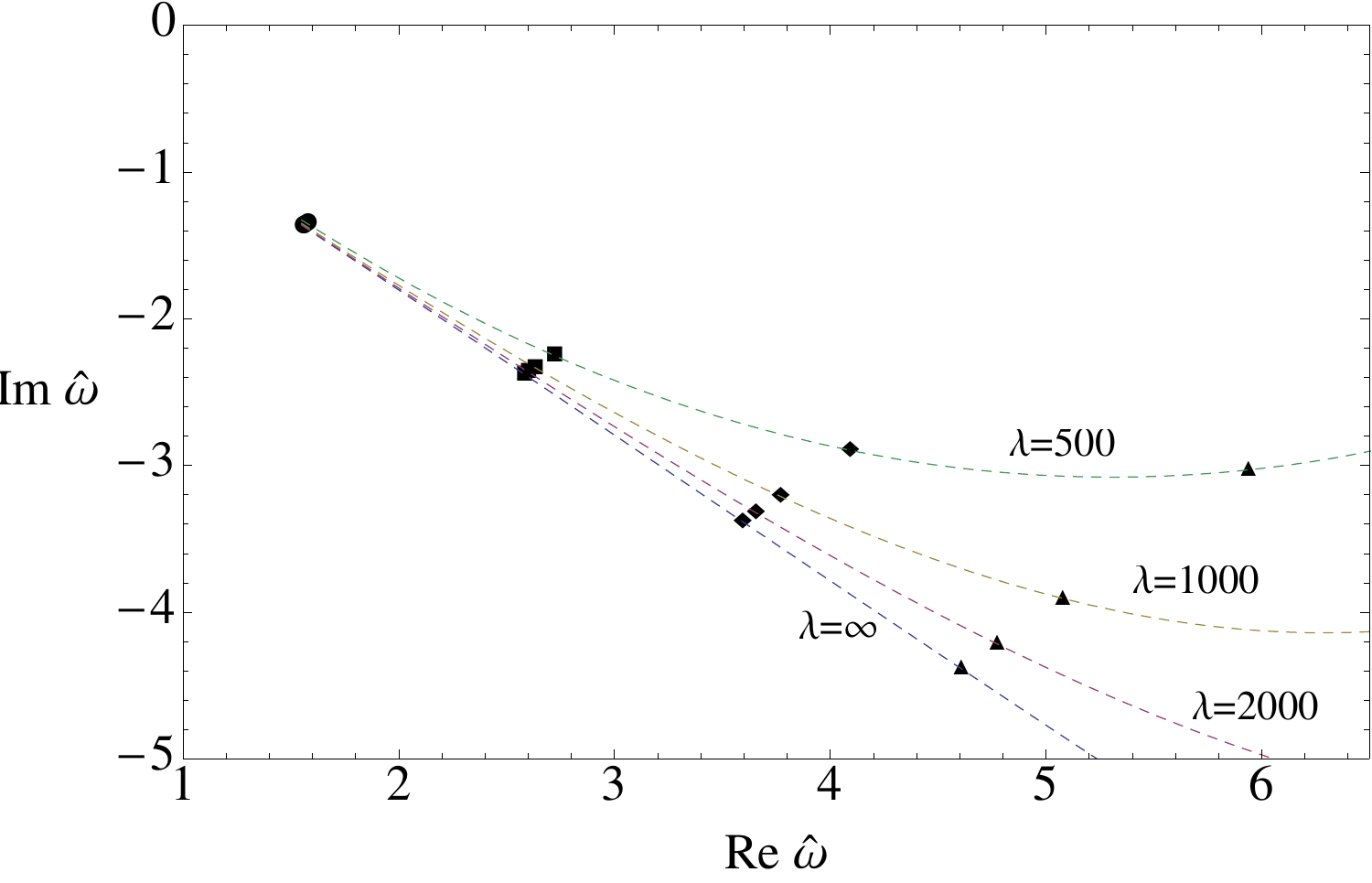}$\;\;\;\;\;\;\;\;$\includegraphics[width=7cm]{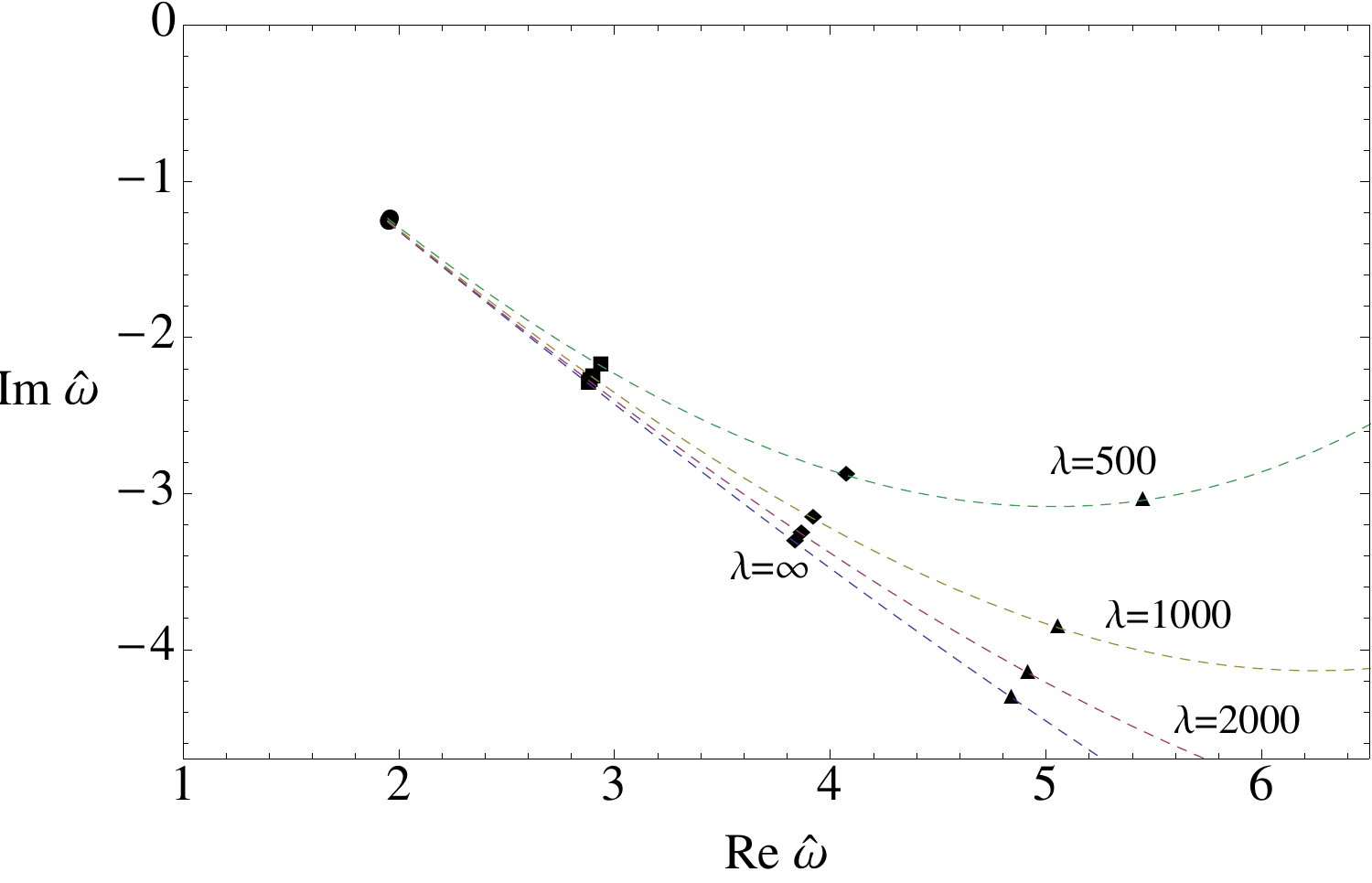}
\caption {The flow of the QNM in the scalar  channel for $q=0$ (left) and $q=2\pi T$ (right) as a function of $\lambda$.  The dashed lines are drawn to illustrate the bending of the tower of QNM towards the real axis as the coupling constant decreases. }
\label{QNM}
\end{figure}

In fig.~\ref{QNM}  we show the flow of the tower of QNM  obtained from the photon and scalar channel correlator. In both cases a clear trend is visible. As the coupling constant is decreased from the infinite coupling limit the imaginary part of $\omega_n$ increases rapidly.  Also note that higher energetic modes show a stronger dependence on the coupling constant, with a larger shift towards the real axis. From equ.(\ref{frobenius}) one deduces that  finite coupling corrections extend  the lifetimes of the excitations.
It should be noted though that the strong coupling expansion can only be trusted when the deviation from the infinite coupling limit is small, which clearly is not the case for all displayed poles.

\subsection{Thermalization of the spectral density}

\begin{figure}
\centering
\begin{minipage}{0.45\textwidth}
\centering
\includegraphics[width=7.1cm]{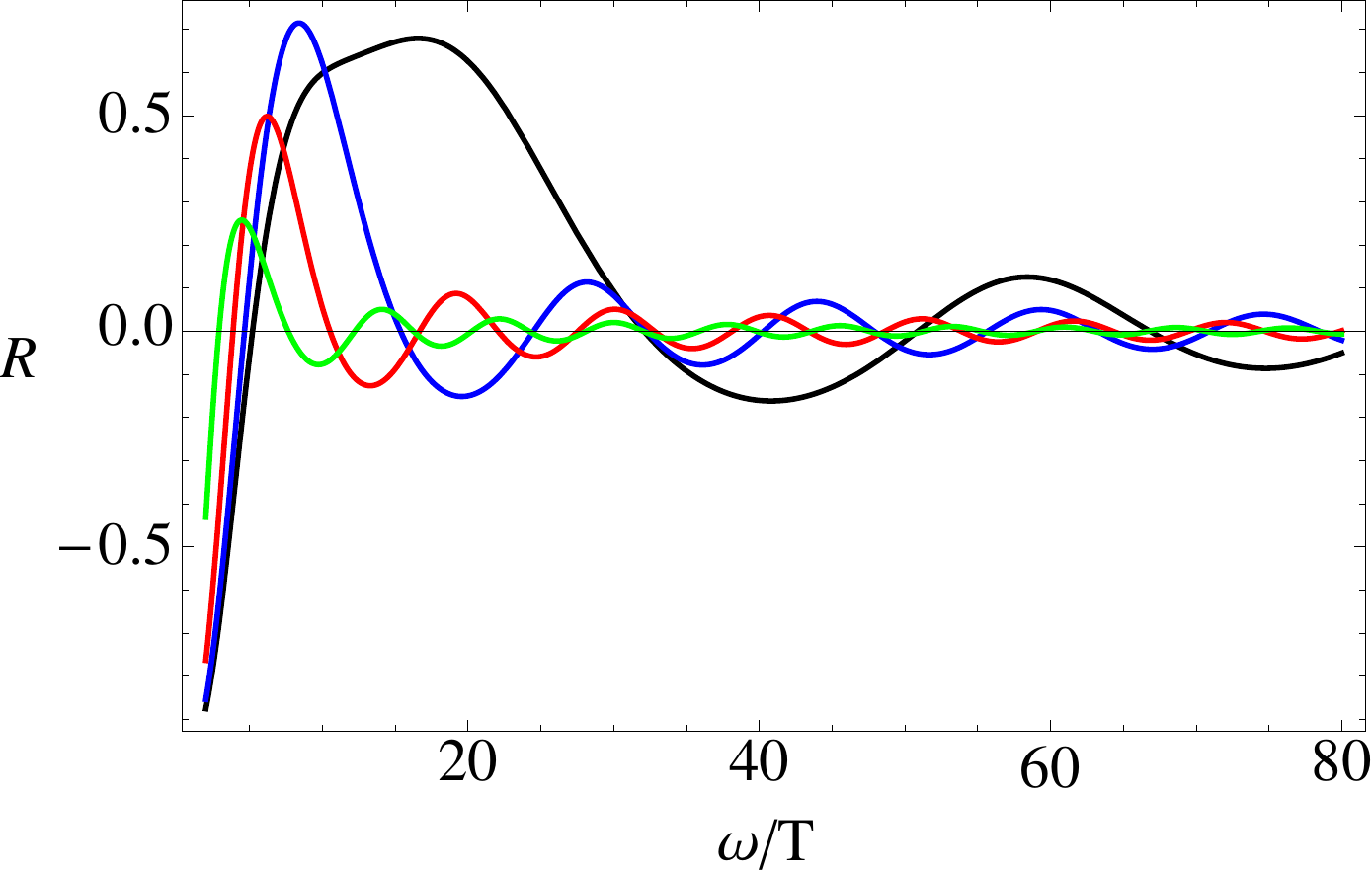}
\caption{R in the scalar channel using $q=0$ and $\lambda=\infty$ for $u_s=0.1,\;0,4,\;0.7,\;0.9$ (from large to small amplitudes).}\label{Rscalar1}
\end{minipage}
\hspace{0.2cm}
\begin{minipage}{0.45\textwidth}
\centering\includegraphics[width=7.1cm]{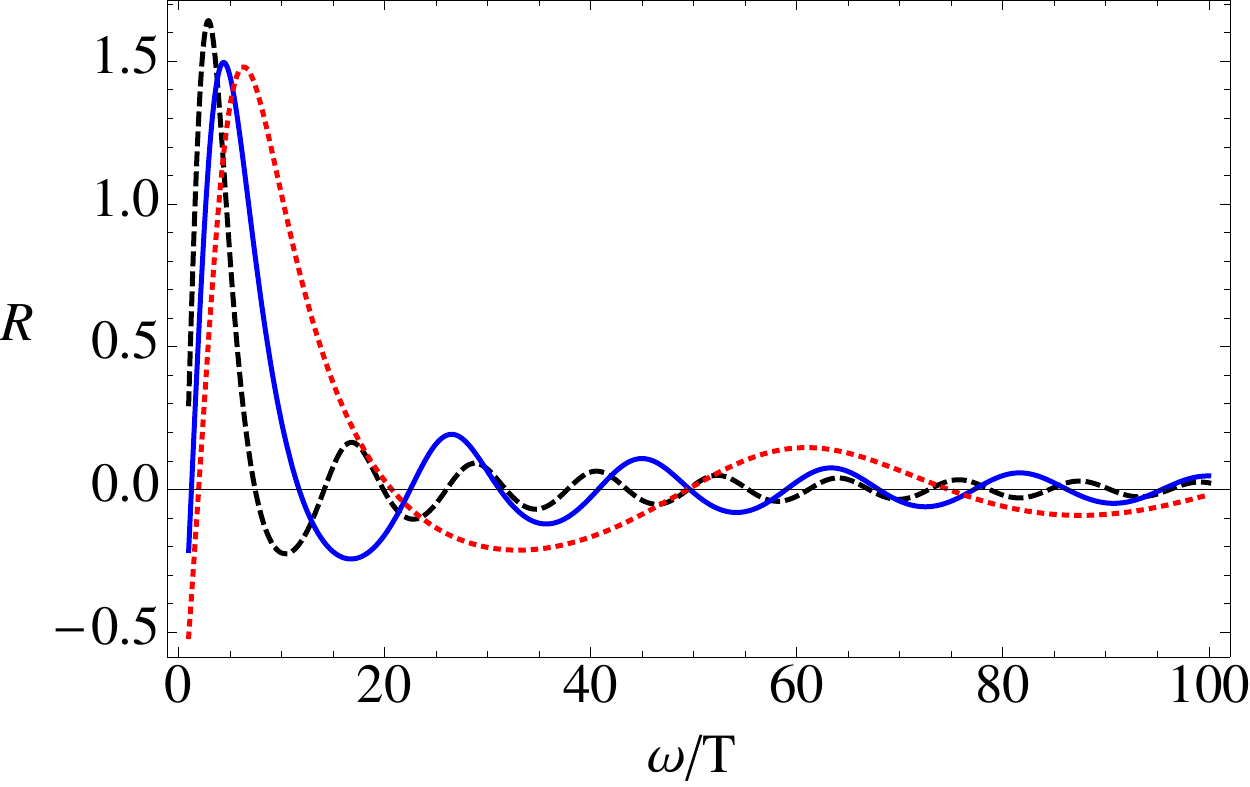}
\caption{R for photons for different virtualities $c=1,\;0.8,\; 0$ (from large to small amplitudes) for $u_s=0.6$.}\label{Rphoton}
\end{minipage}
\end{figure}

In order to study the thermalization pattern of the different plasma constituents it is most instructive to look at the relative deviation of the spectral density from its thermal limit given in (\ref{R}).
Starting with the infinite coupling limit we display in fig.~\ref{Rscalar1} how the relative deviation of the scalar channel behaves for different positions of the shell. As the shell moves towards the horizon the amplitude of $R$ decreases  while the frequency increases, showing that the system thermalizes\footnote{Since the  effects of the shell location are very minor,  in all of our plots we set the parameter of the shell location to rather arbitrary  values which display the effect in the clearest way.}. From this figure one can also see that high energetic modes are closer to equilibrium than the low energetic ones, showing the usual top-down thermalization pattern. The same conclusion holds for all other modes as well, and in the case of photons an  analytic calculation \cite{Steineder:2013ana} reveals  that $R\sim 1/\omega$.

As a next step we look at the virtuality dependence of $R$ in the $\lambda=\infty$ case by letting the virtuality  
$v=(\w^2- q^2)/\w^2$ be a free parameter. We parameterize $q=c\w$ and plot in fig.~\ref{Rphoton}  the relative deviation, $R$, for $c=0$, 0.8 and 1. The most important effect of the virtuality is that the larger the virtuality of the photon is, the smaller is the amplitude of the oscillations. This implies that on shell photons are last to thermalize while maximal virtual photons, i.e. dileptons at rest, thermalize first.

\begin{figure}[h]
\centering
\includegraphics[width=7.1cm]{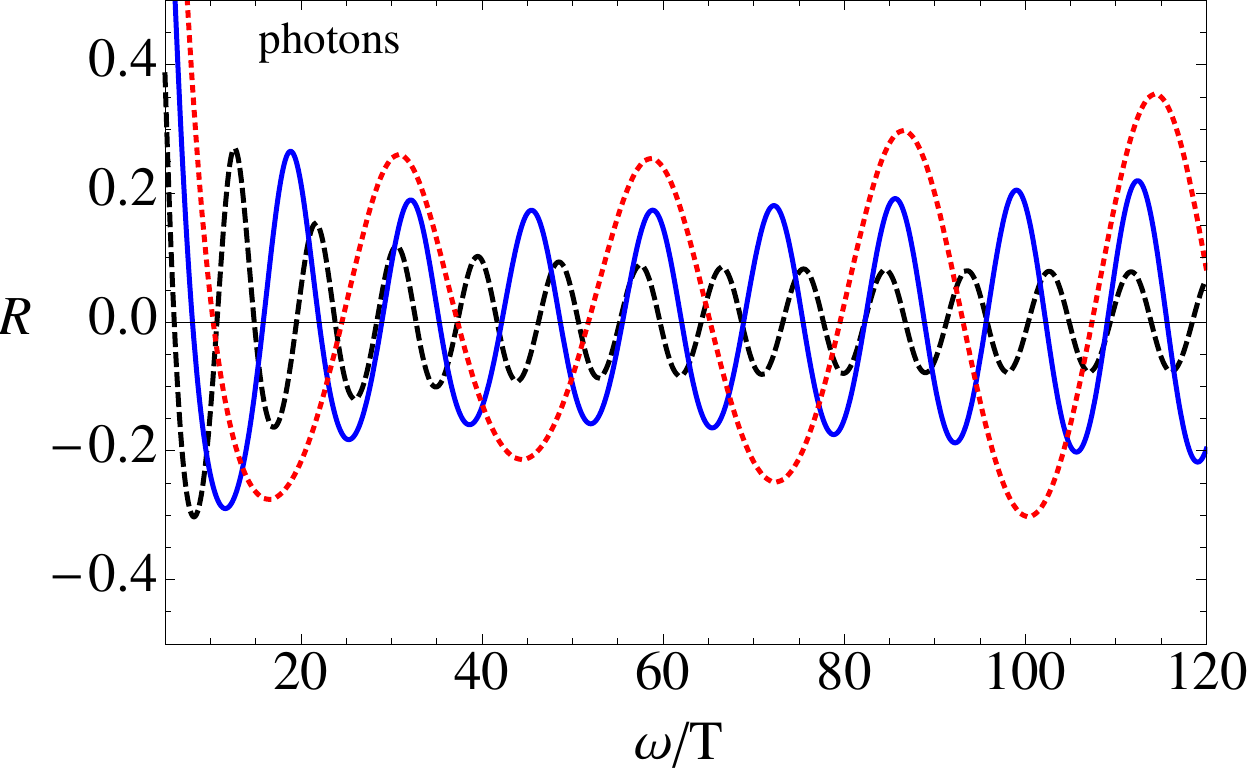}$\;\;\;\;\;\;\;\;$\includegraphics[width=7.1cm]{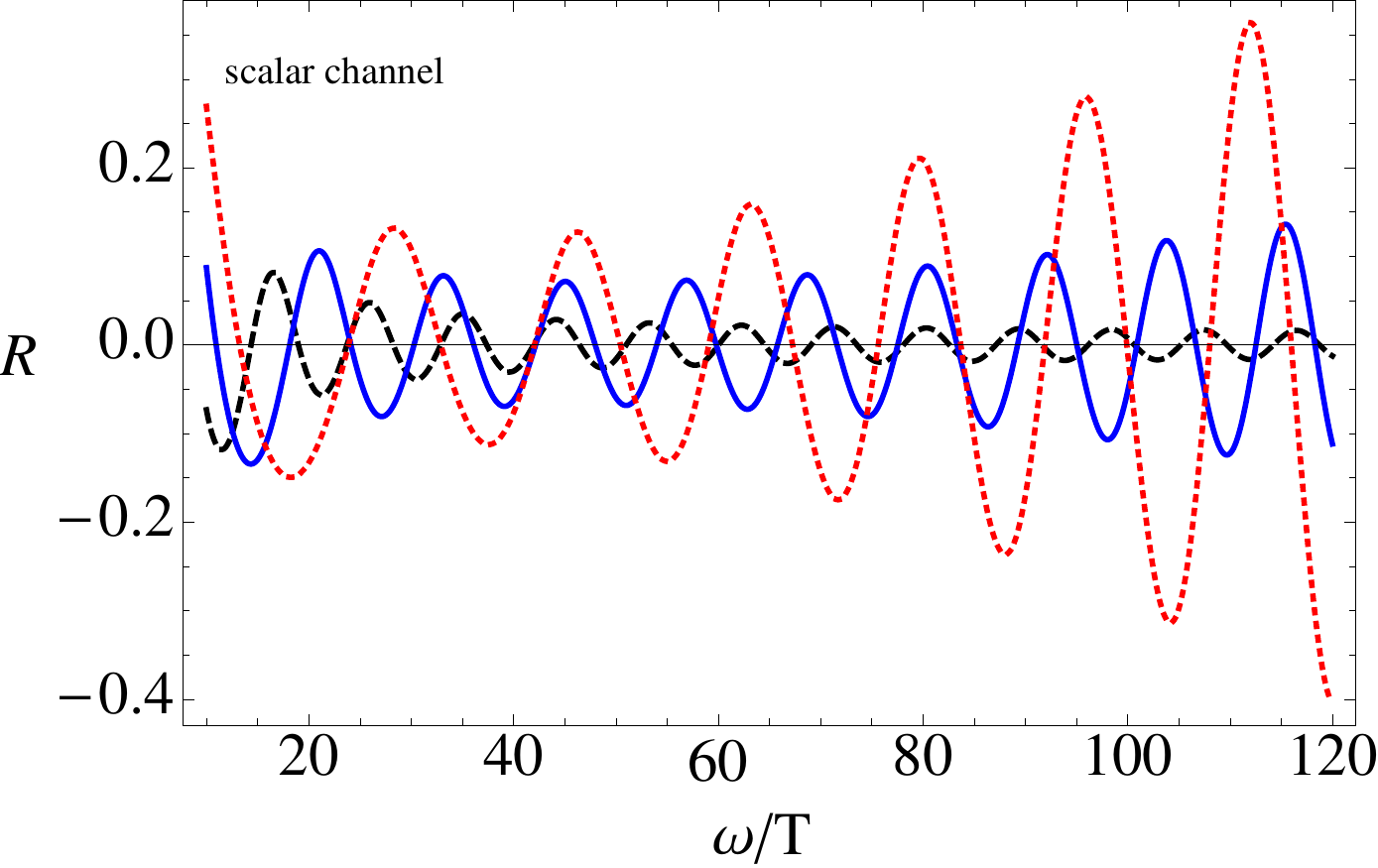}
\caption {Left: The relative deviation $R$ photons for $\lambda=100$ and $u_s=0.8$ and different values of $c=1,\;0.8,\;0$ (from large to small amplitudes). Right: The same values as before in the scalar channel but for $c=8/9,\;6/9,\;0$ (from large to small amplitudes). The rise of R at some critical value points towards a weakening of the usual top down thermalization pattern.
}
\label{R2}
\end{figure}

We are now ready to move on and study  finite coupling effects on the the relative deviation of the spectral density, shown in fig.~\ref{R2}  for photons and the scalar channel for three different values of $c$ at a fixed position of the shell and $\lambda=100$.  For small frequencies the usual top down pattern persists, which however changes for larger values of $\omega/T$.
For $c=0$, the relative deviation approaches a constant. As $c$ is increased $R$ starts to rise again at some critical value. First of all this shows that finite coupling corrections have a larger impact on the high energetic modes, in accordance with the QNM analysis. Second, at large frequencies the relative deviation strongly depends on the value of $c$ (in the case of photons on the virtuality). The larger the value of $c$ the bigger
is the deviation from the thermal limit. We interpret this as a weakening of the usual top-down thermalization pattern moving towards bottom-up at smaller coupling.

\section{Conclusions}

In the paper at hand we studied  holographic  thermalization patterns of $\mathcal{N}=4$ SYM theory and their dependence on the coupling constant.   First, we investigated the flow of the tower of QNM as a function of the 't Hooft coupling. As the coupling decreases from the infinite coupling limit the imaginary part of the quasinormal modes increases, extending the lifetime of the excitation. 
Higher energetic modes show a stronger dependence on the finite coupling corrections. This analysis is particularly useful since it is independent of the thermalization model being used.

Second, we analyzed the relative deviation of spectral densities from their thermal limit  using the collapse of a thin shell of matter and the subsequent formation of a black hole working within the quasi static approximation. At infinite coupling the usual top-down pattern is observed. In the case of photons there is a strong dependence on the virtuality. On shell photons are last to thermalize, whereas dileptons at rest thermalize first.
 
Going away from the infinite coupling limit changes the top-down pattern. At a critical value of the frequency  the relative deviation starts rising again, being further away from its equilibrium value compared to smaller frequencies. 
All observed behaviours point towards that  the usual top-down thermalization pattern at infinite coupling moves towards bottom-up as the coupling constant decreases. 
The effect of finite coupling corrections on thermalizing  geometric probes can be found in \cite{Baron:2013cya}.

\section*{Acknowledgements}

S.S. thanks the organizers of the "EPS-HEP  2013" conference for the invitation and the stimulating atmosphere.  This work was supported by the START project Y435-N16 of the Austrian Science Fund (FWF)

\end{document}